\documentclass{emulateapj}

\usepackage{graphicx}
\usepackage{psfig}
\usepackage{epsfig}
\usepackage{amsmath}

\newcommand{\msun}{\,\hbox{$M_{\odot}$}}

\newcommand{\kms}{\,\hbox{\hbox{km}\,\hbox{s}$^{-1}$}}
\newcommand{\mbh}{\,\hbox{$M_{\rm BH}$}}
\newcommand{\msigma}{\,\hbox{$M_{\rm BH}-\sigma$}}

\newcommand{\lopt}{\,\hbox{$L_{\rm opt}$}}
\newcommand{\spi}{{\it Spitzer}}
\newcommand{\ang}{\,\hbox{\AA}}
\newcommand{\nev}{\,\hbox{[\ion{Ne}{5}]}}
\newcommand{\oiv}{\,\hbox{[\ion{O}{4}]}}
\newcommand{\oiii}{\,\hbox{[\ion{O}{3}]}}
\newcommand{\clii}{\,\hbox{[\ion{Cl}{2}]}}
\newcommand{\feii}{\,\hbox{[\ion{Fe}{2}]}}
\newcommand{\snev}{\,\hbox{$\sigma_{\rm [Ne~\sc{V}]}$}}
\newcommand{\soiv}{\,\hbox{$\sigma_{\rm [O~\sc{IV}]}$}}
\newcommand{\soiii}{\,\hbox{$\sigma_{\rm [O~\sc{III}]}$}}
\newcommand{\xnev}{\,\hbox{$x_{\rm [Ne~\sc{V}]}$}}
\newcommand{\xoiv}{\,\hbox{$x_{\rm [O~\sc{IV}]}$}}
\newcommand{\xopt}{\,\hbox{$x_{\rm opt}$}}
\newcommand{\lnev}{\,\hbox{$L_{\rm [Ne~\sc{V}]}$}}
\newcommand{\loiv}{\,\hbox{$L_{\rm [O~\sc{IV}]}$}}
\newcommand{\lbol}{\,\hbox{$L_{\rm bol}$}}
\newcommand{\lmir}{\,\hbox{$L_{\rm MIR~line}$}}
\newcommand{\eedd}{\,\hbox{$\eta_{\rm Edd}$}}

\shorttitle{High-ionization MIR lines as AGN black hole mass and bolometric luminosity indicators}
\shortauthors{Dasyra et al.}

\begin{document}

\title{High-ionization mid-infrared lines as black hole mass and bolometric luminosity indicators in active 
galactic nuclei}

\author{K. M. Dasyra\altaffilmark{1}, L. C. Ho\altaffilmark{2}, 
L. Armus\altaffilmark{1}, P. Ogle\altaffilmark{1}, G. Helou\altaffilmark{1}, 
B. M. Peterson\altaffilmark{3}, D. Lutz\altaffilmark{4}, 
H. Netzer\altaffilmark{5}, E. Sturm\altaffilmark{4}}

\altaffiltext{1}{Spitzer Science Center, California Institute of Technology,
Mail Code 220-6, 1200 East California Blvd, Pasadena, CA 91125}
\altaffiltext{2}{The Observatories of the Carnegie Institution of Washington,
813 Santa Barbara St., Pasadena, CA 91101}
\altaffiltext{3}{Department of Astronomy, The Ohio State University, 140 West
18th Avenue, Columbus, OH 43210}
\altaffiltext{4}{Max-Planck-Institut f\"ur extraterrestrische Physik,
Postfach 1312, 85741, Garching, Germany}
\altaffiltext{5}{School of Physics and Astronomy, Raymond and Beverly Sackler Faculty of
Exact Sciences, Tel-Aviv University, Tel-Aviv, \hbox{69978, Israel}}

\begin{abstract}
We present relations of the black hole mass and the optical luminosity with the velocity 
dispersion and the luminosity of the \nev\ and the \oiv\ high-ionization lines in the mid-infrared 
(MIR) for 28 reverberation-mapped active galactic nuclei. We used high-resolution \spi\ Infrared 
Spectrograph and {\it Infrared Space Observatory} Short Wavelength Spectrometer data to fit the 
profiles of these MIR emission lines that originate from the narrow-line region of the nucleus. 
We find that the lines are often resolved and that the velocity dispersion of \nev\ and \oiv\ 
follows a relation similar to that between the black hole mass  
and the bulge stellar velocity dispersion found for local galaxies. The luminosity of the \nev\ and 
the \oiv\ lines in these sources is correlated with that of the optical 5100$\ang$ continuum 
and with the black hole mass. Our results provide a means to derive black hole properties
in various types of active galactic nuclei, including highly obscured systems.
\end{abstract}

\keywords{
infrared: galaxies ---
galaxies: active ---
galaxies: nuclei ---
galaxies: Seyferts ---
galaxies: kinematics and dynamics ---
quasars: emission lines
}


\section{Introduction}
\label{sec:intro}
Following the relation between the black hole (BH) mass, \mbh, and the bulge stellar velocity 
dispersion, $\sigma_*$,  (\citealt{ferrarese}; \citealt{gebhardt}; \citealt{tremaine}), a similar 
relation was found for the velocity dispersion $\sigma$ of the \oiii\ 5007\ang\ line (\citealt{nelson00}; 
\citealt{greene}) that originates from the narrow-line region (NLR) gas of the active galactic nucleus (AGN). 
The NLR gas is thought to be mostly gravitationally bound to the bulge in high-luminosity AGNs 
(\citealt{nelson96}; \citealt{greene}; \citealt{laor07}), unlike the broad-line region gas that is 
virialized due its proximity to the BH (\citealt{peterson99}). This relation is useful for systems 
in which the stellar absorption lines cannot be observed, e.g., because of dilution of the stellar 
light by the AGN continuum. In addition to this relation, \mbh\ and the luminosity $L$ 
of the 5100 \ang\ optical continuum, \lopt , are correlated in AGNs with BH measurements (\citealt{kaspi00}). 

There are two main advantages to expanding such relations in the mid-infrared (MIR). The first 
is that the \ion{Ne}{5} and \ion{O}{4} ions emitting at 14.32 and 25.89 \micron\ cannot be easily 
excited by star-forming regions since they have ionization potentials $\chi$ of 97.12 and 54.93 eV. 
The second reason is the low obscuration, which allows for the results to be applied to type 2 
AGNs. In this Letter, we investigate for relations between \mbh\ and \lopt\ with MIR NLR line 
velocity dispersions and luminosities.


\section{The sample and the data reduction}
\label{sec:data}
Our sources were selected from a sample of 35 local AGNs that have \mbh\ measurements from reverberation 
experiments (\citealt{peterson04}). Archival high-resolution \spi\ Infrared Spectrograph (IRS) data are 
available for 28 reverberation-mapped AGNs. Of these, eight were previously observed with the
{\it Infrared Space Observatory} Short Wavelength Spectrometer (SWS) in high resolution (\citealt{sturm02}).

\begin{figure}
\centering
\includegraphics[width=6cm]{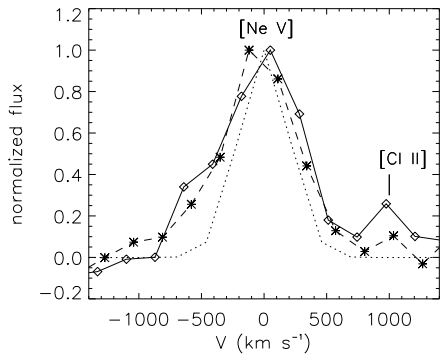}
\caption{\label{fig:ex} Examples of lines resolved with IRS. The \nev\ profile is given for PG\,2130+099 (solid line, 
diamonds) and Mrk\,279 (dashed line, stars). The resolution element is plotted as a dotted line.}
\end{figure}

To reduce the IRS data we first averaged all images at the same nod position, rejecting pixels that 
deviated from the median image by more than 5 times its standard deviation. We computed the uncertainty 
image and then created the bad-pixel mask and extracted the spectrum of each nod position using the 
\spi\ data reduction packages IRSCLEAN and SPICE. For the spectral extraction, we assumed a point-source 
spatial distribution for the high-ionization lines, corresponding to an unresolved NLR. When present, 
we removed the weak \clii\ and \feii\  lines at 14.37 and 25.99 \micron , 1000 and 1200 \kms\ longward 
of \nev\ and \oiv .  We fitted the \nev\ and \oiv\ lines and their underlying continua using Gaussian
and second-order polynomial functions, respectively, which was a good approximation for 
most cases. We considered resolved all lines with $\sigma_m - \epsilon_m > \sigma_i+\epsilon_i$, 
where $\sigma_i$ is the instrumental resolution $R$ at a given wavelength divided by 2.35, 
$\epsilon_i$ is the error of  $\sigma_i$, and $\sigma_{m}$ is the measured velocity dispersion. 
The error of $\sigma_{m}$, $\epsilon_m$,
is computed as $(\epsilon_{std}^2+\epsilon_i^2)^{0.5}$. The quantity $\epsilon_{std}$ is the standard 
deviation of the $\sigma_{m}$ values that are obtained when using different polynomial coefficients
to describe the local continuum. It encapsulates uncertainties in the continuum level and slope. 
The instrumental resolution and its error were measured as a function of wavelength using standard IRS 
calibration targets (P Cygni, HD 190429, HD 174638). 
The average value of $R$ in the 14.0$-$18.0 \micron\ range is 494$\pm$59 \kms. Between 25.0 and 34.2 
\micron, it is equal to 503$\pm$63 \kms. The intrinsic velocity dispersions of the resolved lines 
were computed as $(\sigma_m^2-\sigma_i^2)^{0.5}$ 
and are presented in Table~\ref{tab:data}. Examples of lines resolved with IRS are given in Figure~\ref{fig:ex}. 
To complement the {\it Spitzer} data and to compute the widths of unresolved lines, we used SWS spectra 
presented in \cite{sturm02}. SWS has a higher resolution ($R\sim$160 \kms\ at 14.32 \micron\ and $R\sim$230 
\kms\ at 25.89 \micron) but lower sensitivity than IRS. It detected only the brightest \nev\ and \oiv\ lines, 
the velocity dispersions of which are presented in Table~\ref{tab:data}. To fit the relations between the MIR 
line and the BH properties, we used the FITEXY algorithm that takes into account the error bars in both 
axes (\citealt{press}) and that treats all variables symmetrically (\citealt{tremaine}).

\section{Results}

\subsection{Correlations between \mbh\ and the velocity dispersion of \nev\ and \oiv.} 
\label{sec:mbh_sigma}
\begin{figure*}
\centering
\includegraphics[width=14cm]{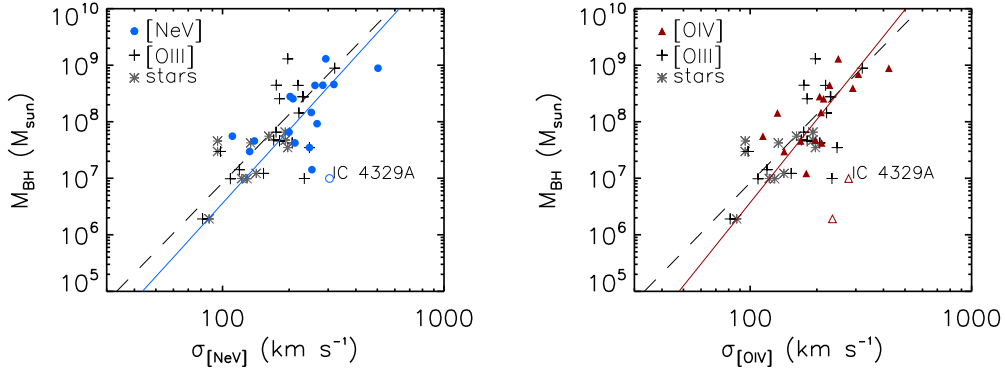}
\caption{\label{fig:ms} The relation between \mbh\ and the velocity dispersion of the MIR NLR lines 
\nev\ (left, circles) and \oiv\ (right, triangles). The solid line corresponds to the best-fit solution
and the open symbols correspond to sources that have been excluded from the fit. For comparison, 
we overplot the \cite{tremaine} relation (dashed line) and the stellar, CaII, velocity dispersion 
measurements of the reverberation-mapped AGNs (stars; \citealt{onken04}; \citealt{nelson04}). The crosses correspond to
$\sigma_{\rm \oiii}$.}
\end{figure*}

The \msigma\ relation of the MIR high-ionization lines is presented in Figure~\ref{fig:ms}. 
Its best-fit solution is
\begin{equation}
\hbox{$\log\left(\frac{\mbh}{\msun}\right)=\alpha_1 + \beta_1\times \log\left(\frac{\sigma}{200{\rm\,km\,s^{-1}} }\right)$},
\end{equation}
with parameter values corresponding to $\alpha_1=7.86\pm 0.09$ and $\beta_1=4.31\pm 0.63$ 
for \nev\ and  $\alpha_1=8.04\pm 0.08$ and $\beta_1=4.89\pm 0.63$ for \oiv. The scatter values 
of the relation along the \mbh\ axis are 0.55 and 0.51 dex for \nev\ and \oiv , respectively. Its 
Spearman rank coefficients are 0.66 and 0.77, assigning probabilities $p$ of 5.0$\times10^{-3}$ 
and  8.3$\times10^{-4}$ to the null hypothesis that the variables are correlated by chance. 
We excluded all sources with \mbh$<10^7$\msun\ from this fit since the IRS resolution is 
insufficient to provide $\sigma$ measurements on the left-hand side of the relation and since
no SWS measurements are available for such \mbh\ values. Moreover, IC4329A is not used in the fit
since its light curves are not indicative of the presence of a BH (\citealt{peterson04}). 
Given that the best-fit slope agrees within the errors with that of the 
\cite{tremaine} formula, the latter can be alternatively used to derive \mbh\ from \snev\ 
and \soiv. If we fix the slope to $\beta_1=4.02$ and solve for $\alpha_1$ using the IDL 
routine MPFIT, we find that $\alpha_1=7.90\pm 0.25$ and $\alpha_1=8.11\pm 0.26$ for \nev\ and \oiv.

\subsection{Correlations between \lopt, \mbh, and the luminosity of \nev\ and \oiv.} 
\label{sec:mir_opt}
\begin{figure*}
\centering
\includegraphics[width=14cm]{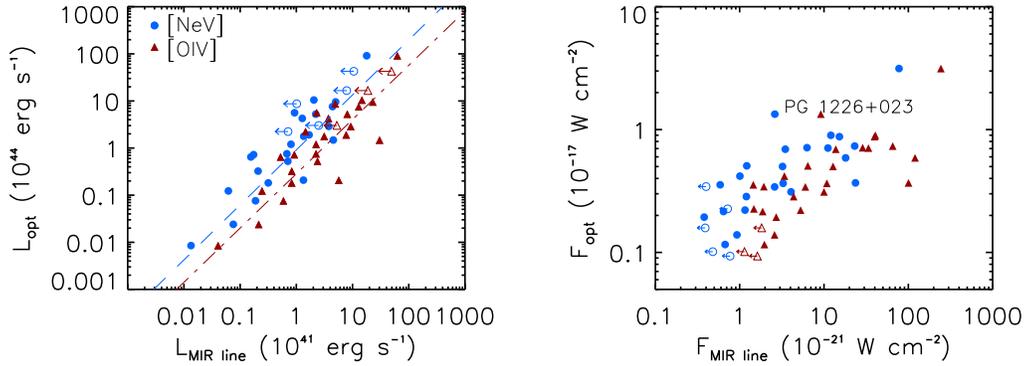}
\caption{\label{fig:ff} The luminosity (left) and flux (right) of the optical continuum scale with 
those of the MIR NLR lines. The best-fit solution to the \nev\ data points (circles) is plotted with a
dashed line and that to the \oiv\ data points (triangles) is plotted with a dash-dotted line. The open
symbols correspond to limits, not used for the fit.}
\end{figure*}

Following the suggestion that the luminosity of \oiii\ can be used as a proxy of \lopt\ 
(\citealt{heckman}), we find that \lnev\ and \loiv\ are related to the 5100$\ang$ continuum 
luminosity (Fig.~\ref{fig:ff}) as
\begin{equation}
\hbox{$\log\left(\frac{\lopt}{10^{44}{\rm \,ergs \,s^{-1}}}\right) =\alpha_2 + \beta_2\times \log\left(\frac{\lmir}{10^{41}{\rm \,ergs\,s^{-1}}}\right)$},
\end{equation}
and they are computed for $H_0$=70 \kms\ Mpc$^{-1}$, $\Omega_{m}$=0.3, and
$\Omega_{\rm total}$=1. The best-fit parameter values 
are $\alpha_2=-0.04\pm 0.01$ and $\beta_2=1.17\pm 0.02$ for \nev\ 
and $\alpha_2=-0.55\pm 0.01$ and $\beta_2=1.15\pm 0.01$ for \oiv. 
The scatter values of the relation are 0.46 and 0.47 dex and its Spearman rank coefficients are 
0.85 and 0.81 for \nev\ and \oiv , respectively, corresponding to probabilities $p$ of 
3.7$\times10^{-7}$ and 7.5$\times10^{-7}$. The relations continue to hold when plotting the 
5100$\ang$ continuum versus the MIR line fluxes (Fig.~\ref{fig:ff}). In the flux-flux diagram, 
the only outlier is PG 1226+023 (3C 273), which is the brightest Palomar-Green QSO (\citealt{schmidt}).

\begin{figure}
\centering
\includegraphics[width=7cm]{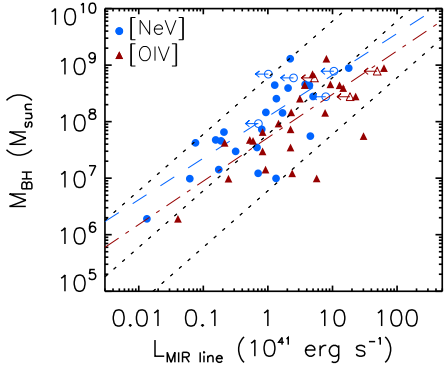}
\caption{\label{fig:ml} The relation between \mbh\ and the luminosity of the MIR NLR lines. The 
symbols are as in Figure~\ref{fig:ff}. The dotted lines correspond to a fixed 
Eddington rate of 0.01, 0.1, and 1 from top to bottom. They were horizontally shifted 
by a factor of 1000 to overlap with the luminosity range of the MIR NLR lines.}
\end{figure}

A correlation similar to that found in the optical between \mbh\ and \lopt\ exists in the MIR between 
\mbh\ and \lnev\ or \loiv\ (Fig.~\ref{fig:ml}). Its best fit is
\begin{equation}
\label{rel3}
\hbox{$\log\left(\frac{\mbh}{\msun}\right)=\alpha_3 + \beta_3\times \log\left(\frac{\lmir}{10^{41}{\rm \,ergs\,s^{-1}} }\right)$},
\end{equation}
with  $\alpha_3=8.09\pm 0.02$ and $\beta_3=0.74\pm 0.03$ for \nev, and
$\alpha_3=7.72\pm 0.02$ and $\beta_3=0.77\pm 0.03$ for \oiv. Its scatter values are 0.46 and 0.50 
dex and its Spearman rank coefficients are 0.76 and 0.69 with $p=2.5\times10^{-5}$ and $p=1.3\times10^{-4}$ 
for \nev\ and \oiv , respectively. We note that upper limits have not been used for the fit.


\section{Discussion: Origin and conditions for the application of the relations }
\label{sec:origin}
The relations between \snev,\soiv\ and \mbh\ indicate that the NLR 
gas kinematics are primarily determined by the potential of the bulge, as it is also 
believed based on the \oiii\ line profiles (\citealt{whittle92}; \citealt{nelson96}; 
\citealt{greene}). However, \snev\ and \soiv\ are on average 67 and 62 \kms\ higher 
than $\sigma_*$, and 51 and 32 \kms\ higher than \soiii . Such discrepancies could be
attributed to an increase of the line width with increasing $\chi$, which is 
often found for optical NLR lines (e.g. \citealt{osterbrock}; \citealt{whittle85b}). It is 
possible that ions with high ionization potentials have high velocity dispersions because they 
are located in NLR clouds that are close to the BH sphere 
of influence. Since the \ion{O}{4} and, mostly, the \ion{Ne}{5} ions are predominantly excited 
by AGNs, the MIR \msigma\ relation can be applied even in systems that undergo starbursts. It 
can also be applied in type 2 AGNs since the lines suffer little from extinction. Moreover,
\snev\ and \soiv\ can be used as surrogates for $\sigma_*$ in environments where measuring 
$\sigma_*$ is hard, such as in bright QSOs. However, the relation does not necessarily hold 
for AGNs with strong winds and jets such as luminous radio sources (\citealt{whittle92}), 
and for AGNs with high Eddington rates, \eedd\ (\citealt{greene}). Such systems can be 
recognized by the asymmetric wings or the high kurtosis of their line profiles (Whittle 
1985a; 1992). 

The luminosities of the broad lines (\citealt{baldwin}) and the narrow lines (Fig.~\ref{fig:ff}) 
scale with \lopt\ and with the bolometric AGN luminosity, \lbol , since they depend on the 
absolute accretion rate onto the central object.
Specifically, \lbol\ is equal to \xopt \lopt, \xnev \lnev, and \xoiv \loiv , where $x$ 
is a wavelength-dependent bolometric correction factor. For \xopt $=$9 (\citealt{kaspi00}; 
\citealt{marconi04}) and for the median values of \lopt/\lnev\ and \lopt/\loiv\ in our sample, 
we find that \xnev $=$13000 and \xoiv $=$4500.  
That the relation does not have a slope of unity could indicate that \xnev\ and \xoiv\ 
depend on $L$ in a manner different than \xopt\ does. \cite{netzer06} found that 
the equivalent width of \oiii\ decreases with increasing \lopt. This behaviour
could be attributed to a different geometric distribution of NLR clouds around the AGN at 
different luminosities. 

Whether the \nev\ or \oiv\ luminosity of a source can be used to derive its \mbh\ partly depends 
on its Eddington rate. To illustrate how changes of \eedd\ affect the relation between the
luminosity of a MIR line and \mbh , we overplot lines of constant \eedd\ 
in Figure~\ref{fig:ml}. The more quiescent a source is, the more its position is shifted to 
the top left corner of this diagram. Equation~(\ref{rel3}) is valid within the \eedd\ range 
0.003$-$0.6 that our sources span, and implies that the mass of a BH determines its 
luminosity output. Its advantage is that it can be applied to sources with obscured optical 
continua or underestimated \oiii\ luminosities, such as type 2 AGNs (\citealt{netzer06}). 
However its scatter is likely to be larger than that presented in \S~\ref{sec:mir_opt} since 
the Eddington rates of the reverberation-mapped AGNs are not necessarily representative of 
those of all local AGNs. 

We conclude that the MIR NLR gas kinematics trace \mbh\ in local 
reverberation-mapped AGNs in a manner similar to stellar kinematics. The \nev\ and \oiv\ 
line widths can be used to estimate \mbh. The calibration of \lnev\ and \loiv\
to \lopt\ provides a new method to compute bolometric luminosities, black hole masses, 
and Eddington rates in various types of AGNs, including highly obscured systems,
albeit with a large scatter.\\
\\
This work was based on observations made with the {\it Spitzer Space Telescope}, 
and was supported by NASA through an award issued by JPL/Caltech.



\evensidemargin=-0.3in
\begin{centering}
\begin{deluxetable}{cccccccccc}
\tablecolumns{10}
\tabletypesize{\tiny}
\tablewidth{0pt}
\tablecaption{\label{tab:data} Black hole and MIR line properties for reverberation-mapped AGNs with high-resolution MIR spectra}
\tablehead{ \colhead{Galaxy} & \colhead{$z$} & \colhead{\mbh } & \colhead{\lopt }
& \colhead{$\sigma_*$} & \colhead{\soiii} 
& \colhead{$F_{[Ne~\sc{V}]} $} & \colhead{\snev} 
& \colhead{$F_{[O~\sc{IV}]} $} & \colhead{\soiv} \\ 
\colhead{} & \colhead{} & \colhead{($10^7$ \msun )} & \colhead{(10$^{44}$ ergs s$^{-1}$) } & \colhead{(km s$^{-1}$) }
& \colhead{(km s$^{-1}$) }  & \colhead{ (10$^{-21}$ W cm$^{-2}$) } 
& \colhead{(km s$^{-1}$) }  & \colhead{ (10$^{-21}$ W cm$^{-2}$) } 
& \colhead{(km s$^{-1}$) } }

\startdata

3C 120      & 0.03301 & 5.55 $\pm$ 2.70 & 1.479 $\pm$ 0.272 & 162 $\pm$ 24 &\nodata & 18.06 $\pm$ 0.23 & 111 $\pm$ 16\tablenotemark{a} & 120.2  $\pm$ 0.8  & 114 $\pm$ 9\tablenotemark{a}\\
Fairall 9   & 0.04702 & 25.5 $\pm$ 5.6  & 1.778 $\pm$ 0.205 & \nodata      & 181    & 2.60 $\pm$  0.33 & 208 $\pm$ 36 & 6.01   $\pm$ 0.28 & 215 $\pm$ 39 \\
IC 4329A    & 0.01605 & 0.99 $\pm$ 1.49 & 0.208 $\pm$ 0.024 & 122 $\pm$ 13 & 234    & 23.63 $\pm$ 1.22 & 304 $\pm$ 44 & 100.6  $\pm$ 1.1  & 278 $\pm$ 30 \\
Mrk 279     & 0.03045 & 3.49 $\pm$ 0.92 & 0.756 $\pm$ 0.087 & 197 $\pm$ 12 & 247    & 3.28 $\pm$  0.23 & 247 $\pm$ 35 & 10.82  $\pm$ 0.29 & \nodata      \\
Mrk 509     & 0.0344  & 14.3 $\pm$ 1.2  & 1.905 $\pm$ 0.351 & \nodata      & 221    & 6.29 $\pm$  0.26 & \nodata      & 28.73  $\pm$ 0.29 & 133 $\pm$ 34\tablenotemark{a} \\
Mrk 590     & 0.02638 & 4.75 $\pm$ 0.74 & 0.646 $\pm$ 0.134 & 192 $\pm$ 10 & 170    & 1.01 $\pm$  0.16 & \nodata      & 3.39   $\pm$ 0.17 & 197 $\pm$ 30 \\
NGC 3227    & 0.00386 & 4.22 $\pm$ 2.14 & 0.024 $\pm$ 0.002 & 134 $\pm$ 6  & 206    & 23.13 $\pm$ 0.35 & 212 $\pm$ 34 & 65.37  $\pm$ 0.59 & 210 $\pm$ 30 \\
NGC 3783    & 0.00973 & 2.98 $\pm$ 0.54 & 0.182 $\pm$ 0.017 & 95 $\pm$ 10  & 98     & 15.24 $\pm$ 1.00 & 133 $\pm$ 21\tablenotemark{a} & 39.78  $\pm$ 0.84 & 142 $\pm$ 25\tablenotemark{a}   \\
NGC 4051    & 0.00234 & 0.191 $\pm$ 0.078 & 0.0085 $\pm$ 0.0010 & 87 $\pm$ 5 & 81     & 11.13 $\pm$ 0.25 & \nodata   & 33.69  $\pm$ 0.52 & 235 $\pm$ 32 \\
NGC 4151    & 0.00332 & 4.57 $\pm$ 0.52 & 0.076 $\pm$ 0.040 & 95 $\pm$ 7  & 181     & 77.64 $\pm$ 1.55 & 139 $\pm$ 26\tablenotemark{a} & 243.5  $\pm$ 1.5  & 169 $\pm$ 13\tablenotemark{a}   \\
NGC 4593    & 0.00900 & 0.98 $\pm$ 0.21 & 0.123 $\pm$ 0.040 & 129 $\pm$ 15 & 109    & 3.48 $\pm$  0.35 & \nodata      & 13.79  $\pm$ 0.86 & \nodata \\
NGC 5548    & 0.01717 & 6.54 $\pm$ 0.26 & 0.324 $\pm$ 0.324 & 192 $\pm$ 15 & 174    & 3.23 $\pm$  0.21 & 200 $\pm$ 33 & 12.80  $\pm$ 0.66 & \nodata      \\
NGC 7469    & 0.01632 & 1.22 $\pm$ 0.14 & 0.525 $\pm$ 0.024 & 142 $\pm$ 3  & 153    & 12.07 $\pm$ 0.63 & \nodata      & 40.63  $\pm$ 3.06 & 179 $\pm$ 22\tablenotemark{a} \\
PG 0003+199 & 0.02578 & 1.42 $\pm$ 0.37 & 0.724 $\pm$ 0.067 & \nodata      & 119    & 1.21 $\pm$  0.30 & 253 $\pm$ 50 & 6.44   $\pm$ 0.28 & \nodata \\
PG 0026+129 & 0.142   & 39.3 $\pm$ 9.6  & 10.47 $\pm$ 1.45  & \nodata      &\nodata & 0.38 $\pm$  0.12 & \nodata      & 2.71   $\pm$ 0.38 & 291 $\pm$ 51 \\
PG 0804+761 & 0.1     & 69.3 $\pm$ 8.3  & 8.710 $\pm$ 1.604 & \nodata      &\nodata & $<0.40$          & \nodata      & 1.93   $\pm$ 0.25 & 308 $\pm$ 57 \\
PG 0844+349 & 0.064   & 9.24 $\pm$ 3.81 & 2.239 $\pm$ 0.206 & \nodata      &\nodata & $<0.72$          & 267 $\pm$ 56 & 1.47   $\pm$ 0.21 & \nodata \\
PG 0953+414 & 0.2341  & 27.6 $\pm$ 5.9  & 16.60 $\pm$ 2.29  & \nodata      & 231    & $<0.48$          & \nodata      & $<1.14$            & \nodata \\
PG 1211+143 & 0.0809  & 14.6 $\pm$ 4.4  & 5.623 $\pm$ 0.777 &\nodata & \nodata      & 0.53 $\pm$  0.14 & 252 $\pm$ 58 & 1.45   $\pm$ 0.26 & 209 $\pm$ 40 \\
PG 1226+023 & 0.1583  & 88.6 $\pm$ 18.7 & 91.20 $\pm$ 10.50 & \nodata      & 321    & 2.61 $\pm$  0.42 & 504 $\pm$ 64 & 9.17   $\pm$ 0.33 & 423 $\pm$ 61 \\
PG 1229+204 & 0.0603  & 7.32 $\pm$ 3.52 & 1.202 $\pm$ 0.138 & 162 $\pm$ 32 &\nodata & 0.93 $\pm$  0.07 & \nodata      & 2.59   $\pm$ 0.45 & \nodata \\ 
PG 1307+085 & 0.155   & 44.0 $\pm$ 12.3 & 7.586 $\pm$ 0.700 & \nodata      & 219    & 0.67 $\pm$  0.06 & 262 $\pm$ 38 & 1.96   $\pm$ 0.33 & \nodata \\
PG 1411+442 & 0.0896  & 44.3 $\pm$ 14.6 & 4.266 $\pm$ 0.393 & \nodata      & 175    & 0.64 $\pm$  0.09 & 284 $\pm$ 54 & 1.87   $\pm$ 0.24 & 228 $\pm$ 52 \\
PG 1426+015 & 0.0865  & 129.8 $\pm$ 38.5 & 5.248 $\pm$ 0.846 & 185 $\pm$ 67 & 197    & 1.20 $\pm$ 0.17 & 292 $\pm$ 46 & 4.37   $\pm$ 0.56 & 250 $\pm$ 38 \\
PG 1613+658 & 0.129   & 27.9 $\pm$ 12.9 & 9.550 $\pm$ 1.099  & \nodata      & 230    & 1.16 $\pm$ 0.21 & 202 $\pm$ 43 & 5.28   $\pm$ 0.25 & 206 $\pm$ 34 \\
PG 1617+175 & 0.11244 & 59.4 $\pm$ 13.8 & 3.020 $\pm$ 0.556 & 183 $\pm$ 47  &\nodata & $<0.77$          & \nodata      & $<1.62$           & \nodata \\
PG 1700+518 & 0.292   & 78.1 $\pm$ 17.4 & 42.66 $\pm$ 2.95  & \nodata       &\nodata & $<0.39$          & \nodata      & $<1.82$           & \nodata \\
PG 2130+099 & 0.063   & 45.7 $\pm$ 5.5  & 2.884 $\pm$ 0.266 & 172 $\pm$ 46 &\nodata  & 4.06 $\pm$ 0.20  & 318 $\pm$ 44 & 10.00  $\pm$ 0.31 & \nodata \\
\enddata
\tablecomments{ The \mbh\ and \lopt\ values are from \cite{peterson04}, Bentz et al.\ (2006; 2007), and \cite{denney06}.
The $\sigma_*$ values were extracted from the CaII triplet (\citealt{onken04}; \citealt{nelson04}) in the Seyferts and from the CO
band heads (\citealt{dasyra07}) in the QSOs. The \soiii\ values are from \cite{peterson04}, Shields et al.\ (2003; 2006),
and \cite{bonning05}. For sources with two measurements of $\sigma$, the average value was used. The MIR line detection limit was set
to 3 times the noise standard deviation.}
\tablenotetext{a}{The velocity dispersions of these lines were extracted from SWS spectra presented in \cite{sturm02}.}
\end{deluxetable}
\end{centering}
\oddsidemargin=0in

\end{document}